%% file: main.tex
\documentclass[runningheads]{llncs}
\usepackage[linesnumbered,ruled,vlined]{algorithm2e}
\usepackage{caption}
\SetKwInput{KwInput}{Input}                
\SetKwInput{KwOutput}{Output}              

\usepackage{chngcntr}
\counterwithin{figure}{section}
\counterwithin{table}{section}

\usepackage{fancyhdr}
\usepackage{tikz}
\usepackage{newfloat}
\DeclareFloatingEnvironment[fileext=dia,within=section]{flowchart}
\usepackage[linesnumbered,ruled,vlined]{algorithm2e}
\usepackage{amsmath}
\usepackage{pgfplots}
\usepackage{textcomp}
\usepackage{dirtytalk}
\usepackage{amsfonts}
\usepackage{xcolor}
\usepackage{tikz}
\usepackage{float}
\usepackage{subcaption}
\usepackage{graphicx}  
\usepackage{fixltx2e}
\usetikzlibrary{shapes,arrows}
\usetikzlibrary{positioning}
\usetikzlibrary{shapes.geometric, arrows}

\tikzset{venn circle/.style={draw,circle,minimum width=3cm,draw=none,fill=#1!40,name path=#1}}
\tikzset{venn circle1/.style={draw,circle,minimum width=3.8cm,draw=none,fill=#1!40,name path=#1}}
\tikzset{venn circle2/.style={draw,circle,minimum width=3cm,draw=none,fill=#1!40,name path=#1}}
\tikzset{venn circle3/.style={draw,circle,minimum width=3.25cm,draw=none,fill=#1!40,name path=#1}}

\usetikzlibrary{intersections}
\usepgfplotslibrary{fillbetween}
\pgfplotsset{compat=1.12}
\usepackage{tikz}
\usepackage{fixltx2e}
\usetikzlibrary{shapes,arrows}
\usetikzlibrary{positioning}
\usetikzlibrary{shapes.geometric, arrows}

\begin{document}
\title{A new Hybrid Lattice Attack on Galbraith's Binary LWE Cryptosystem}
\titlerunning{A new Hybrid Lattice Attack on Galbraith's Binary LWE Cryptosystem}
%
\author{Tikaram Sanyashi  \and
M. Bhargav Sri Venkatesh \and
Kapil Agarwal  \and \\
 Manish Verma  \and
 Bernard Menezes }
%
%
\institute{Indian Institute of Technology, Bombay
\\ \email{\{tikaram,bhargav,kapilagg,manishv,bernard\}@cse.iitb.ac.in}}
%
\maketitle              

\begin{abstract}
    
LWE-based cryptosystems are an attractive alternative to traditional ones in the post-quantum era. To minimize the storage cost of part of its public key - a $256 \times 640$ integer matrix, $\textbf{T}$ - a binary version of $\textbf{T}$ has been proposed. One component of its ciphertext, $\textbf{c}_{1}$ is computed as $\textbf{c}_{1} = \textbf{Tu}$ where $\textbf{u}$ is an ephemeral secret. Knowing $\textbf{u}$, the plaintext can be deduced. Given $\textbf{c}_{1}$ and $\textbf{T}$, Galbraith's challenge is to compute $\textbf{u}$ with existing computing resources in 1 year. Our hybrid approach guesses and removes some bits of the solution vector and maps the problem of solving the resulting sub-instance to the Closest Vector Problem in Lattice Theory. The lattice-based approach reduces the number of bits to be guessed while the initial guess based on LP relaxation reduces the number of subsequent guesses to polynomial rather than exponential in the number of guessed bits. Further enhancements partition the set of guessed bits and use a 2-step application of LP. Given the constraint of processor cores and time, a one-time training algorithm learns the optimal combination of partitions yielding a success rate of 9\% - 23\% with 1000 - 100,000 cores in 1 year. This compares favourably with earlier work that yielded 2\% success with 3000 cores.


\end{abstract}

\input{Introduction}

\input{Background}

\input{Our_approach}

\input{Union}

\input{Related_work}
\input{Conclusion}

\input{References}

\end{document}

%% file: Introduction.tex
\section{Introduction}
\noindent Introduced by Regev  \cite{OR} in 2005, Learning with Errors (LWE) is a problem in machine learning and is as hard to solve as certain worst-case lattice problems. Unlike most widely used cryptographic algorithms, it is known to be invulnerable to quantum computers. It is the basis of many cryptographic constructions including 
IND-CPA and IND-CCA secure encryption \cite{OR} \cite{GPV08}, homomorphic encryption \cite{BV} \cite{BGV},  identity-based encryption \cite{CHKP09} \cite{ABB10}, oblivious transfer protocols \cite{PVW08}, lossy-trapdoor functions \cite{PW11}  and many more.

The LWE cryptosystem performs bit by bit encryption. The private key, $\boldsymbol{s}$, is a vector of length $n$ where each element of $\boldsymbol{s}$ is randomly chosen over $\textbf{Z}_q$, $q$ prime. The corresponding public key has two components. The first is a random $m \times n$ matrix, $\boldsymbol{T}$, with elements  over $\textbf{Z}_q$ and with rows denoted $\boldsymbol{a}_{i}$. The second component is a vector, $\boldsymbol{b}$, of length $m$ where the $i^{th}$ element of $\boldsymbol{b}$ is $\boldsymbol{s}^{T}\boldsymbol{a}_{i} + e_i$ (mod $q$). The $e_i$'s are drawn from a discretized normal distribution with mean 0 and standard deviation $\sigma$.

To encrypt a bit, $x$, a random binary vector (nonce), $\boldsymbol{u}$,  of length $m$ is chosen. This is a per-message ephemeral secret. The ciphertext is ($\boldsymbol{c}_1$, $c_2$)  where $\boldsymbol{c}_{1} = \boldsymbol{T} \boldsymbol{u}$ (mod $q$) and $c_2$ = $\boldsymbol{b} \boldsymbol{u} + x \lfloor q/2 \rfloor $ (mod $q$). A received message is  decrypted to 0 or 1 depending on whether $c_2 - \boldsymbol{c}_1 \boldsymbol{s}$ is closer to 0 or $\lfloor q/2 \rfloor$.

To thwart various lattice-based attacks, Lindner et al. \cite{LP} suggested the values of 256, 640 and 4093 respectively for $n$, $m$ and $q$ leading to a public key of size $640 (256 + 1) log_2 (4093) = 246.7$ Kbytes. However, this is far higher than the size of the RSA or ECC public keys which are less than 1 Kbyte.

 To reduce the key size of original LWE problem, \cite{SDG13} proposed that the matrix $\boldsymbol{T}$ and secret vector $\boldsymbol{s}$ be binary. In addition, possible attacks on the nonce to recover the plaintext were discussed. Galbraith posed two challenges, the first challenge is to compute $\boldsymbol{u}$ given a random $256 \times 400$ binary matrix $\textbf{T}$ and $\boldsymbol{c_{1}}$ = $\boldsymbol{T}\boldsymbol{u}$ in one day using an ordinary PC. The second challenge is to compute $\boldsymbol{u}$ with matrix $\textbf{T}$ dimension $256 \times 640$ in one year using ``current computing facilities". The aim of this work is to address Galbraith's second challenge with a much higher success rate than what was previously achieved \cite{HM},\cite{SNDM}. Ours is a hybrid approach using, both, Linear Programming (LP) and a lattice-based one. 

 In our approach, LP is initially applied on the given instance. Based on the LP output, an initial guess of $p$ bits of the secret is made. A sub-instance is created after removing the $p$ guessed bits. By using a lattice-based approach (rather than  Integer Linear Programming (ILP) \cite{SNDM}) to solve the reduced instance, we require $20\%$ fewer bits of the solution vector to be guessed and removed. This results in a $300\%$ saving in computation time. The problem of obtaining the remaining bits of the secret is mapped to the Closest Vector Problem (CVP) in lattice theory. If the initial guess is unsuccessful, a fresh guess is made and the process is repeated until we run out of compute power or time.
 The substantial decrease in execution time is due to, both, a fewer number of bits to be guessed and, further, a considerable reduction in the expected number of guesses from exponential to polynomial in $p$ due to the application of LP.
 
 The second enhancement involves a two-step application of LP. After its first application, $p$ bits are guessed and removed. LP is again applied on the reduced sub-instance after which $230-p$ bits are guessed and removed. The lattice-based approach is then employed on the doubly reduced sub-instance. The key advantage of of the two-step approach is that the expected number of errors in the $230-p$ guessed bits is reduced compared to the case with a single application of LP. This translates to a substantial reduction in the number of guesses (and hence execution time).
 
 For a given $p$ and the initial guesses of the $p$ and $230-p$ bits (based on the LP output), we hypothesize that there are respectively $e_1$ and $e_2$ errors in those bits. Different hypotheses are defined by varying the values of $p$, $e_1$, and $e_2$. Our strategy is to learn which hypotheses are satisfied by the largest number of the instances while simultaneously factoring the computation time to process the guesses (the number of guess is a function of $p, e_1,e_2$). In the training phase, we use a variant of the Budgeted Maximum Coverage Problem to iteratively select the best set of hypotheses subject to optimizing an objective function that incorporates the number of new instances added and the execution time. By combining multiple hypotheses, we obtain a substantial improvement in success rate or equivalently a considerable reduction in execution time for a given success rate.
 
 The paper is organized as follows. Section 2 presents background material. Our contributions are presented as three strategies - the first two are presented in Section 3 and the third is in Section 4. Section 5 is a brief summary of related work.  Section 6 concludes the paper.
 
 
 

\begin{table}[h!] 
\centering
\caption{Notations}
\begin{tabular} {|c|c|} 
 \hline
 Symbol & Meaning\\
 \hline
 $\textbf{u}_s$ & LP output vector after sorting and round-off \\
 \hline
 $\textbf{u}_{s}^{'}$ & Sorted, rounded-off $640-p$ bit vector after applying LP twice \\
 \hline
 $\textbf{u}_{s_1}, \textbf{u}_{s_2} , \textbf{u}_{s_3}$ & First $230$, next $256$ and last $154$ bits of solution vector $\textbf{u}_{s}$\\
 \hline
 $p$ & Number of bits removed after first application of LP\\
 \hline
 $e_{1}$ & Number of errors in $p$ bits.\\
 \hline
 $e_{2}$ & Number of errors in next $230-p$ bits.\\
 \hline
 $t_{LP}$ & Computing time of Linear Programing (LP) \\
 \hline
 $t_{LR}$ & Computing time of Lattice Reduction (LR)\\
 \hline
 $t_{B}$ & Computing time of Babai's Nearest Plane algorithm \\
 \hline
 $s(p, e_{1}, e_{2})$ & Instances with $e_{1}$ errors in first $p$ bits and $e_{2}$ errors in $230-p$ bits.\\
 \hline
 $t(p, e_{1}, e_{2})$ & Total time to compute Algorithm 1 with input parameters $p, e_{1}, e_{2}$ \\
 \hline
\end{tabular}

\label{tab:multipleregimeILP}
\end{table}

%% file: Background.tex
\section{Background}

A Lattice $L$ is a discrete additive subgroup of $\mathbb R^{n}$. Equivalently, $L$ is comprised of integer linear combinations of a set of linearly independent vectors. A lattice is represented by a basis, a set of $m$ linearly independent integer vectors $(b_{1},b_{2}, \cdots ,b_{m})$ each of size $n$ which generates the lattice 
\begin{equation*}
 L(b_{1},b_{2}, \cdots ,b_{m})=\bigg \{ \mathlarger{\mathlarger{\sum}}_{i=0}^{m} x_{i} b_{i} \ : \ x_{i}\in \mathbb Z \bigg \}
\end{equation*}
A lattice can have multiple bases and 
a basis is usually represented by $m \times n$ matrix where $m$ basis vectors are rows of the matrix. 

One of the hard problems in lattice based cryptography is the Closest Vector Problem (CVP).
The problem is to find a lattice vector $u$ given a basis $B$ of some lattice $L=L(B)$ and a non-lattice vector $v \in \mathbb R^{n}$ with minimum $||u-v||$. One of the best known solutions to solving CVP is the Nearest Plane algorithm developed by Babai\cite{BNP}.



Before applying Babai's Algorithm, lattice reduction is performed to obtain short and near-orthogonal basis vectors. Some of the different lattice reduction algorithms are LLL, BKZ and BKZ2.0.
BKZ \cite{SE} algorithms behave differently based on block size $k$. 
In practice, run time of BKZ increases rapidly with block size and becomes practically infeasible for $k > 30$ or so. 
Chen and Nguyen \cite{CN} presented an updated version of BKZ i.e. BKZ 2.0. 
It uses extreme pruning techniques of Gama-Nguyen-Regev \cite{GNR}
 that significantly decreases the running time of enumeration subroutine without degrading its output quality allowing much higher block size ($\beta \geq 40$) in high dimension.  

Implementations of LLL, BKZ  and BKZ 2.0 are available in many software packages, notably in NTL\cite{VS}, FLINT\cite{HJP} and fplll\cite{CPS}. 
In our implementation, we have used BKZ 2.0 from fplll library with block size $k=22$.

For pre-processing and guessing some bits of the ephemeral secret, u we use Linear Programming (LP).
LP is used to determine the best possible solution from a given list of requirements represented in the form of linear relationships. 

The standard algorithm for solving LP is 
the Simplex Algorithm though it is not guaranteed to run in polynomial time. Later, 
it was shown that LP could be done in polynomial time by using the Ellipsoid Algorithm (but it tends to be fairly slow in practice). Karmarkar proposed a much faster polynomial-time algorithm - the first of a class of so-called “interior-point” methods.

In our implementation, we have used Matlab's inbuilt linear programming function, linprog.
There are three variants of linear programming algorithms available in Matlab 
viz dual-simplex, interior-point (default) and interior-point-legacy. Often, the dual-simplex and interior-point algorithms are fast and use least memory. The interior-point-legacy method is similar to the interior-point algorithm but uses more memory and is slower and less robust. 
We have used the default algorithm provided by Matlab to obtain the initial guess of u.



%% file: Our_approach.tex
\section{Our Approach}
We introduce Strategy 1 which involves a lattice-based approach in conjunction with LP. Strategy 2 which involves a 2-step application of LP is introduced next. We begin by reviewing Strategy 0 which uses ILP.
\subsection{Strategies 0, 1 and 2}
As stated earlier, we attempt to obtain $u$ given $\textbf{T}$ and $\textbf{c}_{1}$ in the equation below
\begin{equation}
    \textbf{Tu} = \textbf{c}_{1} \label{master}
\end{equation}

Our approach is summarized in Flowchart \ref{dia:flowchart}. The pre-processing step creates an approximate solution to Equation \ref{master}. We then guess selected bits in $\textbf{u}$ (Step 1) and create a reduced sub-instance (Step 2) by removing  the guessed bits in $\textbf{u}$, deleting the corresponding columns of matrix $\textbf{T}$ and re-computing the value of $\textbf{c}_{1}$. Formally, if the values of the guessed bits are $u^{(i_1)}, u^{(i_2)}, \dots ,u^{(i_r)}$ in positions $i_1, i_2,\ \dots ,i_r$, then those bits are removed from  $\textbf{u}$, the $i_{1}^{th}, i_{2}^{th},\ \dots ,i_{r}^{th}$ columns in 
$\textbf{T}$ are removed and new value of $\textbf{c}_{1}$ is computed as

\begin{equation} \label{ReduceProblem}
\textbf{c}^{'}_{1} = \textbf{c}_{1} -  \sum_{i = i_1,i_2,...,i_r} u^{(i)} T^{(i)}
\end{equation} 

\indent We then solve the resulting sub-instance and verify (Step 3) whether the computed solution is binary and satisfies Equation \ref{master}. If not, we proceed with the remaining guesses until the secret is obtained or we run out of guesses or resources. While the approach is straightforward, several issues need to be addressed in its implementation.\\
1. Which bits of $\textbf{u}$ may be guessed and what are the values of the guessed bits?\\
2. What method/technique should be used to solve the sub-instance (Step 2)?\\
3. If all the guessed bits are correctly guessed, then will the desired solution be obtained?\\

\indent We next outline several implementation strategies - Strategy 0 was adopted in \cite{SNDM} while strategies 1-3 are newly introduced here.\\
\begin{flowchart}[!ht]
\begin{center}
\tikzstyle{decision} = [diamond, draw, fill=blue!10,text width=6em, text badly centered, inner sep=0pt]
\tikzstyle{block} = [rectangle, draw, fill=blue!10, text width=15em, text centered, minimum height=2em]
\tikzstyle{Startstop} = [rectangle, draw, fill=red!20, text width=4em, text centered, rounded corners, minimum height=2em]
\tikzstyle{mycircle} = [circle, thick, draw=orange, minimum height=4mm]
\tikzstyle{arrow} = [thick,->,>=stealth]
\tikzstyle{line} = [draw, -latex']
\begin{tikzpicture}[align=center,node distance = 1.5cm, font=\small, auto]
    \node [block]                                         (Preprocess) {Pre-process the given instance};
    \node [decision, below of=Preprocess,node distance = 2.1cm]                       (Decision1)  {Guesses exhausted ?};
    \node [block, below of=Decision1,node distance = 2.1cm ]                    (Make) {\raggedleft 1) Make next guess of some bits of solution vector };
    \node [block, below of=Make,node distance = 1.1cm]  (Create)     {\raggedleft 2) Create and solve sub-instance};
    \node [block, below of=Create,node distance = 1.1cm]                         (Verify)     {\raggedleft 3) Verify solution };
    \node [decision, below of=Verify,node distance = 2cm]  (Correct)    {Is solution correct ?};
    \node [Startstop, below of=Correct,node distance = 2cm]                   (Success)    {Success};
    \node [Startstop, right of=Success,node distance = 3cm]                   (Fail)       {Failure};
    \draw [arrow] (Preprocess)          --          (Decision1);
    \draw [arrow] (Decision1)                --          (Make);
    \draw [arrow] (Make)           --          (Create);
    \draw [arrow] (Create)               --          (Verify);
    \draw [arrow] (Verify)              --          (Correct);
    \draw [arrow] (Correct)             -- node [near start] {Yes}                                               (Success);
    \draw [arrow] (Decision1)           -|              (3,-10)            node [near start]{Yes}              (Fail);
    \draw [arrow] (Correct)             -- node [near start] {No}           (-3,-8.35)          |-                (Decision1);

\end{tikzpicture}

\caption{Flowchart expounding the approach in Strategies 0 and 1}
\label{dia:flowchart}
\end{center}
\end{flowchart}

\textbf{Strategy 0:} The LP formulation below from \cite{HM} is employed to solve Equation \ref{master} (Step 0).\\

\begin{center}
\hspace{-20mm} Optimization function: $F(u) = 0$\\
Constraints: $\textbf{Tu}$ = $\textbf{c}_{1}$,  $0 \leq u_i \leq   1 ,\ 1 \leq i \leq m$\\
\end{center}

\indent For the size of matrix $\textbf{T}$ under consideration, the elements of the solution vector are fractions between 0 and 1. These are sorted in order of increasing proximity to 0.5 and then rounded to 0 or 1. Let $\textbf{u}_s$ denote the resulting vector and let $\textbf{T}_s$ be the matrix obtained by re-arranging columns of $\textbf{T}$ in the same order in which the bits of $\textbf{u}$ are re-arranged to obtain $\textbf{u}_s$ so that
\begin{equation}
    \textbf{T}_s \textbf{u}_s = \textbf{c}_{1} \label{sortmaster}
\end{equation}
Extensive experiments conducted in \cite{SNDM} indicate that the bits in $\textbf{u}_s$ differ from the corresponding bits in $\textbf{u}$ in roughly 20\% of the positions. Moreover, the probability that a bit in $\textbf{u}_s$ is in error increases with its position (from left to right). To corroborate those findings, we generated 10,000 random instances and performed LP on each. We found that the average error probability in the first 100 bits is 0.02 while it is 0.44 for the last 100 bits. \\

\indent \cite{SNDM} report that if all errors in the first 280 bits of $\textbf{u}_s$ were corrected, then the sub-instance created after removing these bits could be solved using Integer Linear Programming (ILP) with almost 100\% success rate. The first 280 bits in $\textbf{u}_s$ is our initial guess. Subsequent guesses are obtained by flipping different combinations of at most $e$ of those 280 bits. Thus, there are ${\sum}_{i=0}^{e} {280 \choose i}\ $ guesses to be made. For each guess, we create a reduced sub-instance and solve it using ILP. \\

In our experiments, we found that less than 12\% of the instances have 11 or fewer errors in the first 280 bits of $\textbf{u}_s$. The number of guesses for the partial secret is hence  ${\sum}_{i=0}^{11} {280 \choose i}\ \sim 1.8 \times 10^{19}$. Processing a guess involves an ILP computation. The computation time for ILP is input-dependent. \cite{SNDM} placed a limit of 30 seconds on an ILP instance. Hence, the total time to process all guesses is about a billion years using $3000$ cores for a success rate of 12\%. We next outline a strategy wherein it suffices to guess only 230 bits rather than 280 bits.\\

\textbf{Strategy 1:} As in Strategy 0, we compute $\textbf{u}_s$ and create a smaller sub-instance by removing 230 bits in $\textbf{u}_s$. Unlike Strategy 0, we use a lattice-based approach to solve the resulting sub-instance. Based on extensive experiments with 10,000 instances, we found that it suffices to remove only the first 230 bits from $\textbf{u}_s$ to guarantee a solution for the sub-instance with probability $\sim 1$.

\indent For a given guess of the first $230$ bits of $\textbf{u}_s$, the remaining bits of the secret are computed as follows. Let ($\textbf{T}_{s1},\textbf{T}_{s2},\textbf{T}_{s3}$) be a partitioning of $\textbf{T}_s$. Here $\textbf{T}_{s1}$, $\textbf{T}_{s2}$ and $\textbf{T}_{s3}$ are $256 \times 230$, $256 \times 256$ and $256 \times 154$ sub-matrices. Let $\textbf{u}_s =(\textbf{u}_{s1}, \textbf{u}_{s2},\textbf{u}_{s3})$ be the corresponding partitioning of $\textbf{u}_s$ into sub-vectors of length $230$, $256$ and $154$ respectively. Using Equation \ref{sortmaster}, we have 

\begin{equation} 
\textbf{T}_{s1} \textbf{u}_{s1} +  \textbf{T}_{s2} \textbf{u}_{s2} + \textbf{T}_{s3} \textbf{u}_{s3} = \textbf{c}_{1} \label{eq4}
\end{equation}
Re-arranging and pre-multiplying by $\textbf{T}_{s2}^{-1}$,
\begin{equation}
\textbf{T}_{s2}^{-1} \textbf{T}_{s3} \textbf{u}_{s3}  \equiv  - \textbf{u}_{s2} + \textbf{T}_{s2}^{-1}(\textbf{c}_{1} - \textbf{T}_{s1} \textbf{u}_{s1}) \,\,mod\, \alpha
\label{eq5}
\end{equation}
$\textbf{T}_{s2}^{-1}$ will, in general, have fractional values. So, we compute $\textbf{T}_{s2}^{-1}$ and all terms of Equation \ref{eq5} modulo a large prime, $\alpha$ \footnote{We experimented with different sizes and values of $\alpha$. The selected value of bit size is a trade-off between success probability and lattice reduction time. 26-bit primes yielded a success rate of 99\%.}. $\textbf{u}_{s2}$ is a binary vector and is of negligible norm compared to $\textbf{T}_{s2}^{-1}(\textbf{c}_{1}-\textbf{T}_{s1}\textbf{u}_{s1})$. So
\begin{equation}
(\textbf{T}_{s2}^{-1} \textbf{T}_{s3}) \textbf{u}_{s3}  \approx  \textbf{T}_{s2}^{-1}(\textbf{c}_{1} - \textbf{T}_{s1} \textbf{u}_{s1}) \label{eq6}
\end{equation}

\indent The LHS of Equation \ref{eq6} is a vector in the lattice with basis $\textbf{T}_{s2}^{-1}\textbf{T}_{s3}$ while the RHS is a non-lattice vector. Hence the problem of computing $\textbf{u}_{s3}$ maps to the classical Closest Vector Problem (CVP) in the theory of lattices.

Assuming $e$ of the first $230$ bits of $u_{s}$ are in error, the time to discover the secret is 
\begin{equation}
    t_{LP} + t_{LR} + \sum_{i=0}^{e} {p \choose i}t_{B}
\end{equation}

\textbf{Strategy 2:} As in Strategy 1, we guess $230$ bits of the solution vector but we do so in two steps. Algorithm 1 summarizes the procedure.

\begin{algorithm}[H]
\caption {Hybrid Lattice Approach}
\label{heuristic}

\KwInput{$\textbf{T}_{256\times 640}$, $\textbf{c}_{256 \times 1}$, $p$, $e_1$ and $e_2$}
\KwOutput{$\textbf{u}$ (with significant probability $\textbf{Tu}=\textbf{c}_{1}$)}
  $\textbf{function:} \textsc{ourLP(\textbf{X}, \textbf{z})}$\\
   $\hspace{1.5em}$ Create an LP instance for $\textbf{Xy}=\textbf{z}$  \\ 
    \hspace{1.5em} Use the LP solver to obtain $\textbf{y}$ \\
    \hspace{1.5em} Sort $\textbf{y}$, round it and call it $\textbf{y}_{s}$ \\ 
    \hspace{1.5em} \tcp{Sorting is in order of increasing proximity to 0.5}
    \hspace{1.5em} Arrange columns of $\textbf{X}$ in the same order in which elements of $\textbf{y}$ are  \hspace{2em}  arranged to obtain $\textbf{y}_{s}$ and call it $\textbf{X}_{s}$.\\
 $\textbf{return} (\textbf{X}_{s}, \textbf{y}_{s})$ \\

\BlankLine
\BlankLine

 $(\textbf{T}_{s}, \textbf{u}_{s}) = \textsc{ourLP} (\textbf{T}, \textbf{c}_{1})$ \\

\tcp{Let $\textbf{T}_{s_{2}}$ and $\textbf{T}_{s_{3}}$ respectively denote columns $231-486$ and $487-640$ of $\textbf{T}_{s}$.}
 Compute $\textbf{T}_{s_{2}}^{-1}$ mod $\alpha$ and perform lattice reduction on $\textbf{T}_{s_{2}}^{-1} \textbf{T}_{s_{3}}$\\

 $\textbf{v}_{1} = $first $p$ bits of $\textbf{u}_{s}$\\
\For{each vector $\textbf{v}^{'}_{1}$ such that $\left \| \textbf{v}^{'}_{1} - \textbf{v}_{1} \right \|_1 \leq e_{1}$}
{
     $\textbf{c}^{'}$ = $\textbf{c}_{1} - \sum_{i=1}^{p} v_{1}^{'(i)} T_{s}^{(i)}$ \\
     $\textbf{T}^{'}_{256 \times (640-p)}$ consists of last $(640-p)$ columns of $\textbf{T}_{s}$\\
     $(\textbf{T}^{'}_{s}, \textbf{u}^{'}_{s})$ = \textsc{ourLP}$(\textbf{T}', \textbf{c}')$ \\
     $\textbf{v}_{2} =$ First $(230-p)$ bits of $\textbf{u}'_{s}$\\
    \For{each vector $\textbf{v}^{'}_{2}$ such that $\left \| \textbf{v}^{'}_{2} - \textbf{v}_{2} \right \|_1 \leq e_{2}$}
    {
         Substitute $\textbf{v}_{1}^{'}$ and $\textbf{v}_{2}^{'}$ for $\textbf{u}_{s_1}$  in Equation \ref{eq6}\\
         Solve CVP using Babai's Algorithm\\
        \If{correct solution is obtained}
        {
             Output $\textbf{v}^{'}_{1}, \textbf{v}^{'}_{2}$
        }    
    }
}
Output -1
\end{algorithm}

The function, $\textsc{ourLP}$ is invoked to create a sorted, rounded-off solution vector, $\textbf{u}_s$ as in Strategy 1. The first $p$ bits of $\textbf{u}_s$ is the initial guess of these bits. In each iteration of the outer loop, a fresh guess is made by flipping some combination of $e_{1}$ or fewer of those $p$ bits. The $p$ bits are removed and a sub-instance of size $640-p$ is created.  $\textsc{ourLP}$ is invoked to obtain $\textbf{u}_{s}^{'}$, a partial solution vector of size $640-p$. Each iteration of the inner loop involves guessing the first $230-p$ bits of $\textbf{u}_{s}^{'}$ by flipping a different combination of $e_{2}$ bits. Thus, a total of $p+(230-p)$= $230$ bits are guessed to obtain $\textbf{u}_{s1}$ in Equation \ref{eq6}

A CVP instance is created and solved to obtain $u_{s3}$. This procedure continues until it runs out of guesses or resources. Based on the above description of Algorithm 1, it is clear that LP and Lattice Reduction (LR) are applied once per iteration of the outer loop while Babai's algorithm is executed once per iteration of the inner loop. The total time to run Algorithm 1 is thus  

\begin{equation}
t(p,e_1,e_2) = t_{LP} + \sum_{i=0}^{e_1} {p \choose i} \Big[ t_{LP} + t_{LR} + \sum_{j=0}^{e_2} {230-p \choose j} \times t_{B} \Big] \label{time}
\end{equation}.

The notations and values of the execution times for the various operations are listed in Table \ref{RunTimeDifferentAlgo}. The times were measured on Intel i5 Gen 4, with 3.5 GHz clock and 8 GB DRAM running Ubuntu 16.04 64-bit LTS. The LP solver of Matlab 2015b was used. BKZ with block size=22 implemented in Sage and Babai's Nearest Plane algorithm were used.   

\begin{table}[h!] 
\centering
\caption{Time taken by different algorithm's }
\begin{tabular}{ |c|c|c|  }

 \hline
 Algorithm & Implemented In & Time (seconds) \\
 \hline
 Linear Programming & Matlab & $t_{LP}$ = 0.5\\
 \hline
 Int. Linear Programming & Matlab & 30 (bound)\\
 \hline
 Babai's NP Algo. & Sage & $t_{B}$ = 8\\
 \hline
 BKZ with $\beta=22$ & Sage & $t_{LR}$ = 10500\\
 \hline
\end{tabular}
\label{RunTimeDifferentAlgo}
\end{table}

\subsection{Results}
To estimate success probability, we created a training set of $n=10,000$ randomly generated instances. LP was applied on each instance, the LP output was sorted and rounded. The first $p$ bits of the resulting solution vector ($\textbf{u}_{s}$) were compared with the corresponding bits of the actual secret to determine $e_{1}$, the number of bits in error. The $p$ bits of $\textbf{u}_{s}$ were corrected and removed to create a sub-instance over which LP was again applied. The reduced solution vector was sorted and rounded to obtain $\textbf{u}_{s}^{'}$. The first $230-p$ bits of $\textbf{u}_{s}^{'}$ were compared with the corresponding bits of the true secret to determine the number of bits in error, $e_{2}$. The instance was then added to the ``instance set'', $s(p, e_{1}, e_{2})$ - this is the set of instances with $e_{1}$ errors in the first $p$ bits of $\textbf{u}_{s}$ and $e_{2}$ errors in the first $230-p$ bits of $\textbf{u}_{s}^{'}$. This was carried out for all $n = $ 10,000 instances.  $\frac{|s(p, e_{1}, e_{2})|}{n}$ is a reasonable estimate of the success probability of running Algorithm 1 with input parameters $p$, $e_{1}$ and $e_{2}$. 

The computation times and success probabilities were computed for varying $e_{1}$ and $e_{2}$ and for $p$ ranging from $0$ to $230$ in steps of $5$. The total computation time assumes the values in Table \ref{RunTimeDifferentAlgo} and the availability of 3000 cores. The value of $p$ which maximizes success probability is shown in Table \ref{PartitionTable}. For a fixed value of $e_{1} + e_{2}$, this value of $p$ increases with $e_{1}$. 

\begin{figure}[H]
\centering
\begin{subfigure}[b]{\textwidth}
  \includegraphics[width=1\linewidth]{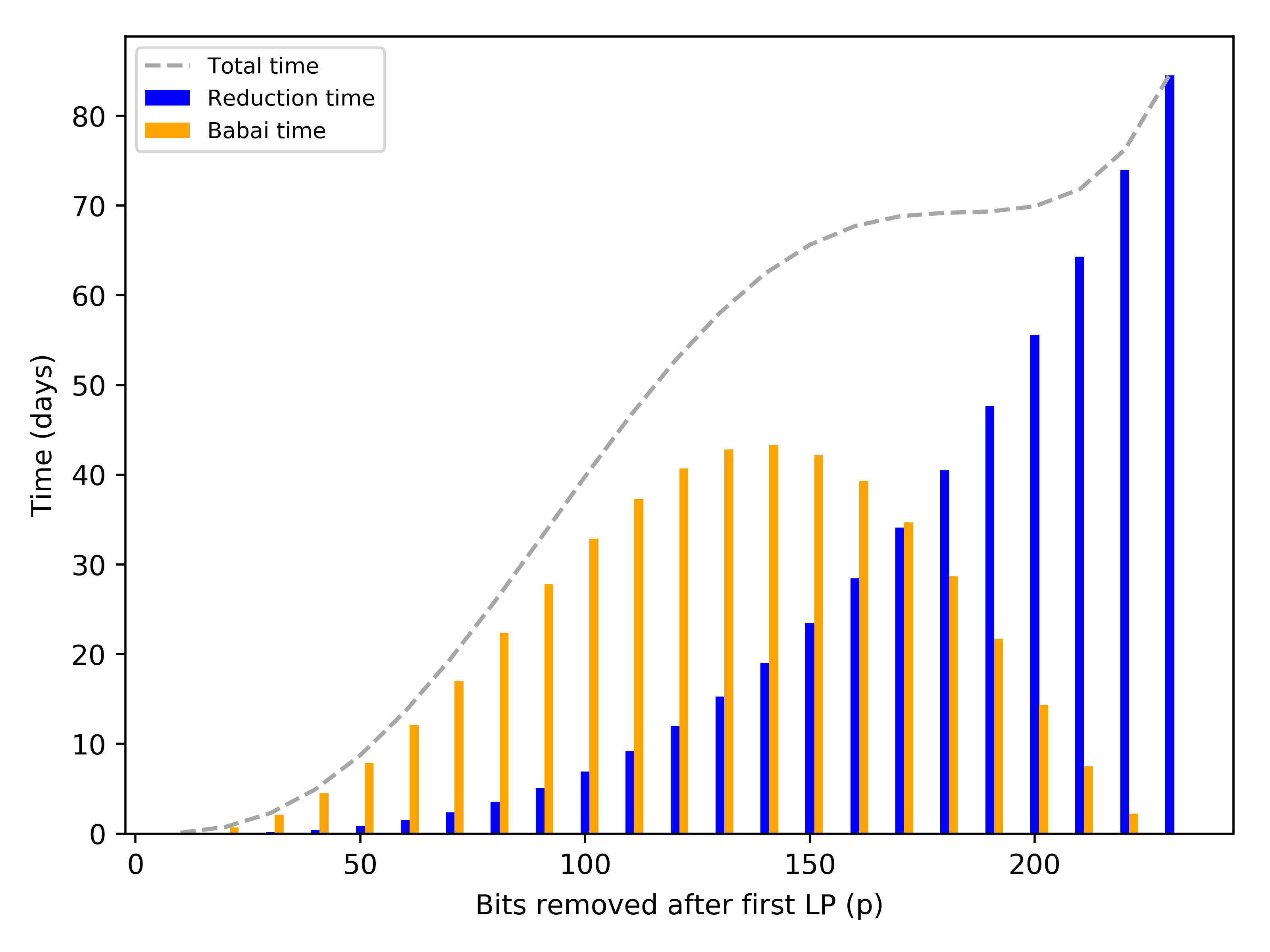}
  \caption{Components of Execution Time for $(e_1, e_2) = (3, 2)$}
  \label{Total_time} 
\end{subfigure}

\begin{subfigure}[b]{\textwidth}
  \includegraphics[width=1\linewidth]{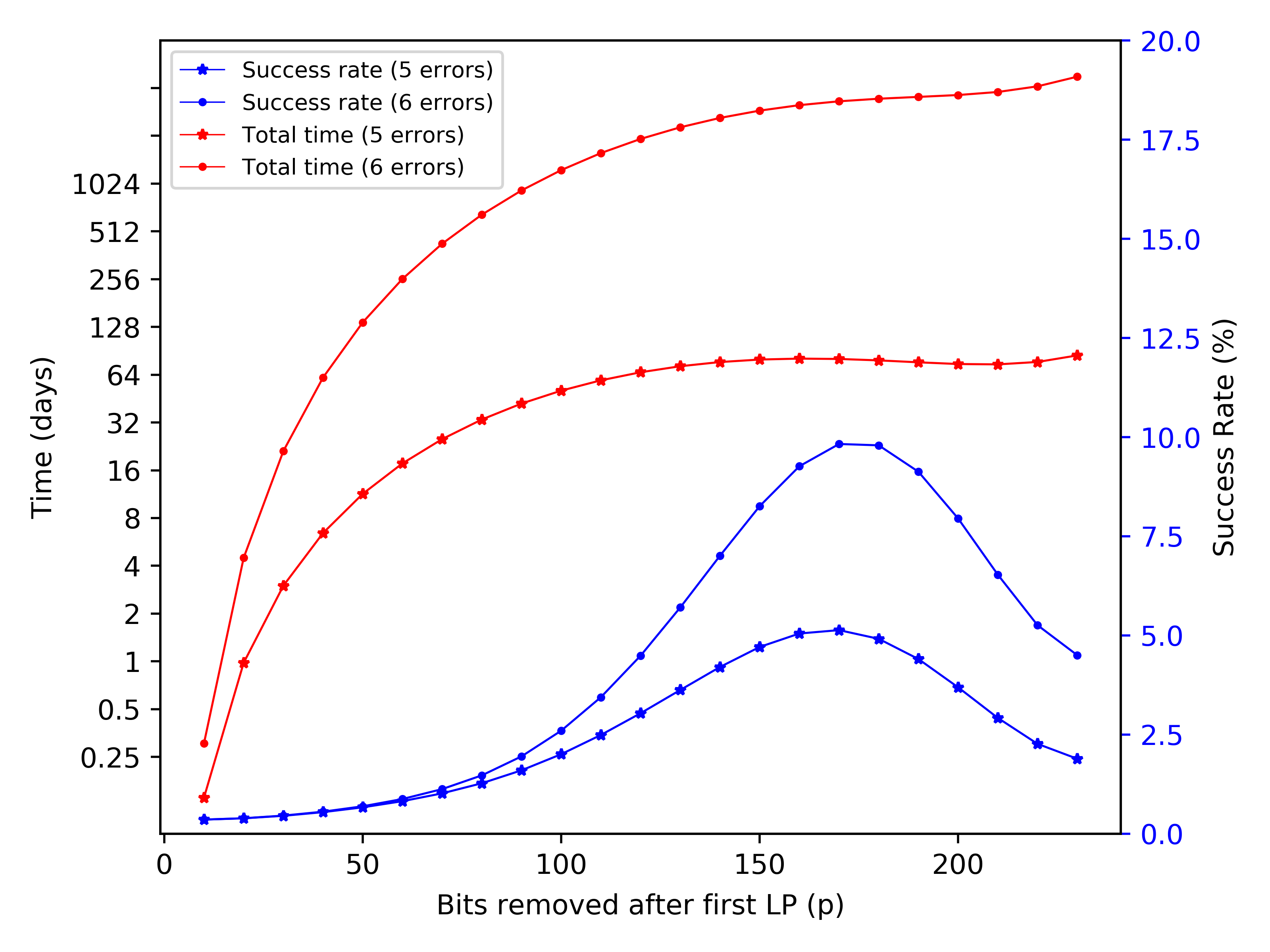}
  \caption{Comparison of $e_1 + e_2 = 5$ $(e_1=3,e_2=2)$ and $e_1 + e_2 = 6$ $(e_1=4,e_2=2)$}
  \label{Time_Success}
\end{subfigure}

\caption{Success Probability and Computation Time as a function of $p$ }
\end{figure}

\begin{table}[h!] 
\centering
\caption{Value of $p$ which maximizes success probability for given (e1, e2) pair with 3000 cores}
\begin{tabular}{ |c|c|c|c| }
 \hline
 $e_{1}$, $e_{2}$ & $p$ & Success Probability & Time (days)\\
 \hline
 0, 5 & 0 & 5\% & 162 \\
 \hline
 1, 4 & 75 & 4\% & 56 \\
 \hline
 2, 3 & 150 & 4.7\% & 30 \\
 \hline
 3, 2 & 175 & 6.1\% & 80 \\
 \hline
 4, 1 & 195 & 6.4\% & 2,551 \\
 \hline
 0, 6 & 0 & 9.4\% & 6,105 \\
 \hline
 1, 5 & 75 & 7.7\% & 1,694 \\
 \hline
 2, 4 & 130 & 8.2\% & 1,075 \\
 \hline
 3, 3 & 160 & 10.6\% & 1,234 \\
\hline
 4, 2 & 175 & 11.8\% & 3,448 \\
\hline
 5, 1 & 195 & 12.4\% & 97,999 \\
 \hline
\end{tabular}
\\
\label{PartitionTable}
\end{table}

The maximum value of success probability increases with $e_{1}$ (beyond $e_{1}$=0). This is at the cost of sharply escalating computation times (beyond $e_{1}$=1). The superiority of Strategy 2 over Strategy 1 is also on display -  $e_{1}$=0 corresponds to Strategy 1. Note that Strategy 2 provides a higher success rate of $6.1\%$ (for $p=175, e_{1}=3, e_{2}=2$ ) versus $5\%$ (Strategy 1). Moreover the computation time of the former is only $80$ days compared to $162$ days for the latter. 

\begin{figure}[H]
\includegraphics[width=12cm, height=7cm]{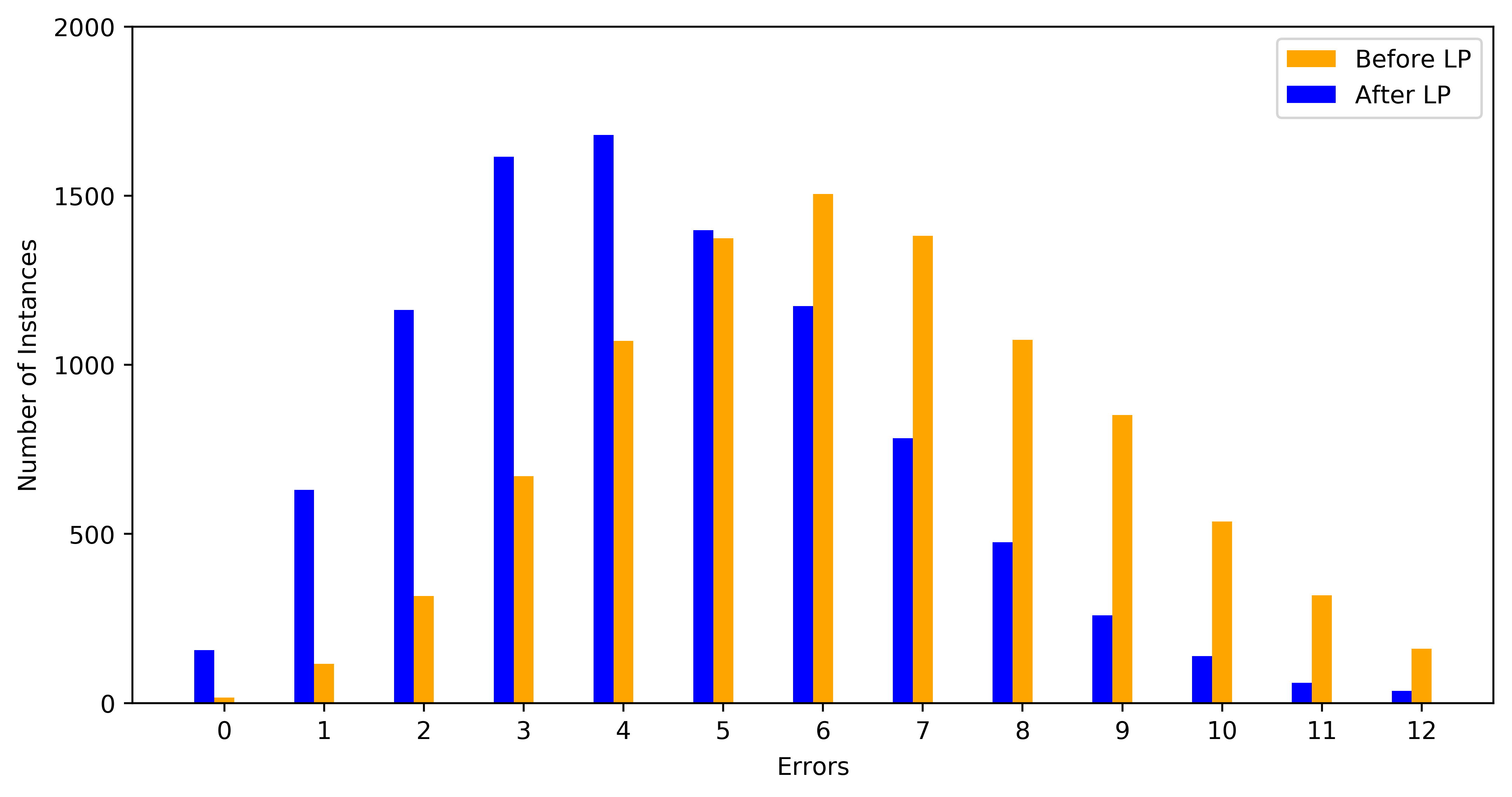}
\centering
\caption{Distribution of errors before and after second application of LP}
\label{Before_After_LP}
\end{figure}

The variation of the execution time of Algorithm 1 denoted $t(p, e_1, e_2)$ with $p$, $e_{1}$ and $e_{2}$ can be better understood by examining the contribution of its various components. The computation time is dominated by the execution of the Babai's Algorithm and Lattice Reduction (LR) (to a first-order approximation the time for Linear Programming may be ignored). 

LR is executed $\sum_{i=0}^{e_1} {p \choose i}$ times while Babai's Algorithm is executed 
$\sum_{i=0}^{e_1} {p \choose i} \times \sum_{j=0}^{e_2} {230-p \choose j}$ times.
As shown in Figure \ref{Total_time}, the contribution of the latter to $t(p, 3, 2)$ peaks at $p$ $\approx$ 140 while that of the former increases with $p$. Overall, $t(p, 3, 2)$ increases monotonically. For $e_{1}>3$, LR dominates while for $e_{1}<3$, Babai's algorithm dominates the computation time. 

With Strategy 1, only a single execution of the compute-intensive LR operation is performed. However, the number of executions of Babai's algorithm, $\sum_{i=0}^{e} {230 \choose i}$ is significantly higher than that with Strategy 2. For example, to achieve success probabilities of $5\%$ and $6\%$ with Strategies 1 and 2, the number of executions of Babai's algorithm are respectively $5.2 \times 10^{9}$ and $1.4 \times 10^{9}$ resulting in an execution time of $162$ and $42.5$ days with 3000 cores respectively. Thus, even though the time spent executing LR with Strategy 2 is $37.2$ days, the overall time of Strategy 2 is less than $50\%$ that of Strategy 1. 

The difference in the number of executions of Babai's algorithm in Strategies 1 and 2 is partially explained by examining the distribution of errors in the first $230-p$ bits of $\textbf{u}_{s}^{'}$ before and after the second application of LP. The two distributions (Figure \ref{Before_After_LP}) have a similar shape but the latter is shifted left. Hence,the second application of LP reduces the errors which in turn necessitates fewer guesses and iterations of the inner loop of Algorithm 1. 


Solving instances with $6$ errors in the first $230$ bits ($e_{1} + e_{2} = 6$)  greatly increases the success probability but at the expense of vastly higher execution time (Figure \ref{Time_Success}). The latter is because the number of guesses (and hence execution time) is exponential in the number of errors. Once again, Strategy 2 yields a much higher success rate ($11.8\%$ versus $9.4\%$) with only $60\%$ of the execution time required by Strategy 1 (Table \ref{PartitionTable}). The maximum success rate is $12.4\%$ but at the cost of 3000 cores running continuously for about 300 years! In the next section, we unveil a strategy which achieves much higher success rate but executes in only 1 year.

%% file: Union.tex
\section{Strategy 3}


Strategy 2 attempted to run Algorithm 1 with the best possible parameter values ($p$,  $e_1$ and $e_2$) determined from experiments on a training set of 10,000 instances. The highest success probability obtained was $6.1\%$ with $3000$ cores in a year.  

Our next strategy (Strategy 3) is to greatly improve on this success rate by running Algorithm 1 repeatedly with different parameter values subject to resource constraints. To illustrate this idea, consider the three parameter sets in  Table \ref{diff_part}. If Algorithm 1 is run in isolation with each of the parameter sets shown, the success rates are $3.9\%$, $4.2\%$ and $5.9\%$ with $3000$ cores in $52$, $50$ and $80$ days respectively. However if Algorithm 1 is run twice with the first two parameter sets, a total of $|s(80, 1, 4) \cup s(120, 2, 3)|= 616$ instances are likely to succeed.

\begin{table}[h!] 
\caption{Performance of Algorithm 1 with single and multiple parameter sets}
\centering
\begin{tabular}{|c|c|c|c|}
 \hline
 $(p,e_1,e_2)$ & $\mid S(p,e_1,e_2) \mid$ & $t(p,e_1,e_2)$ \\
 \hline
(80,1,4) & 391 & 52\\
 \hline
(120,2,3) & 422 & 50\\
 \hline
(170,3,2) & 593 & 80\\
 \hline
Union & 900 & 182\\
 \hline
\end{tabular}
\label{diff_part}
\end{table}

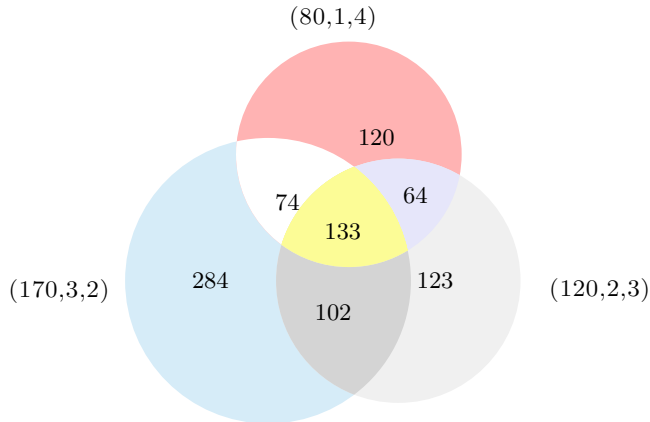
\begin{figure}[!ht]
\centering
\begin{tikzpicture}
\definecolor{green}{rgb}{.6,0.81,0.93}
\definecolor{blue}{rgb}{1,.25,.25}
\definecolor{red}{rgb}{.85,.85,.85}
\definecolor{antiflashwhite}{rgb}{0.95, 0.95, 0.96}
\definecolor{azure}{rgb}{0.94, 1.0, 1.0}
\definecolor{lavender}{rgb}{0.9, 0.9, 0.98}
\definecolor{lightgray}{rgb}{0.83, 0.83, 0.83}
\definecolor{yellow}{rgb}{0.99, 0.99, 0.59}

\node[font=\scshape] at (90:3cm) {(80,1,4)};
    \node [venn circle1=green,label={[label distance=-1.5cm]180: 284 }] (A) at (-150:1cm) {};
    
\node[font=\scshape] at (190:3.7cm) {(170,3,2)};
    \node [venn circle2=blue,label={[label distance=-1.5cm]20: 120 }] (B) at (80:1.2cm) {};
    
\node[font=\scshape] at (-190:-3.6cm) {(120,2,3)};
    \node [venn circle3=red, label={[label distance=-1.5cm]0: 123 }] (C) at (-30:1cm) {};
    
\fill[white,
          intersection segments={
            of=green and blue,
            sequence={R2--L2}
          }]; 
\fill[lightgray,
          intersection segments={
            of=red and green,
            sequence={R1--L2--R0}
          }];
\fill[lavender,
          intersection segments={
            of=red and blue,
            sequence={R2--L2}
          }];
\path [
    name path=rag,
    intersection segments={
        of=green and red,
    }];
\fill[yellow,intersection segments={of=blue and rag,sequence=R2--B1}]  
    [intersection segments={of=rag and blue, sequence={--R2}}];               
    \node[font=\footnotesize,left,yshift=2mm] at (barycentric cs:A=1/2,B=1/2 ) {74}; 
    \node[font=\footnotesize,below,yshift=-2mm] at (barycentric cs:A=1/2,C=1/2 ) {102};  
    \node[font=\footnotesize,right,xshift=2.8mm,yshift=3mm] at (barycentric cs:B=1/2,C=1/2 ) {64}; 
    \node[font=\footnotesize,right,xshift=-3mm,yshift=1mm] at (barycentric cs:A=1/3,B=1/3,C=1/3 ) {133}; 
\end{tikzpicture}
\caption{Union of different instance sets appearing in Table \ref{diff_part}}
\label{union}
\end{figure}

If Algorithm 1 is run a third time with the parameter set $(170,3,2)$, the overall success rate increases to $9\%$. This is achieved with 3000 cores running continuously for about $6$ months. The increases in success rates are best visualized with the Venn diagram in Figure \ref{union}.   

To maximize success probability, it is necessary to identify the parameter sets with which Algorithm 1 should be run so that the union of the corresponding instance sets is maximized while constraining the total execution time to $1$ year. However, this is not straightforward given that total number of instance sets is $276$ (since $p$ is varied from $5$ to $230$ in steps of $5$ and $e_{1}+e_{2} \leq 5$). Our problem maps to the ``Budgeted Maximum Coverage Problem" known to be NP-hard. \cite{SAJ} proposes a greedy heuristic for the above problem and shows that their solution is within $(1-\frac{1}{\epsilon})$ of the optimal solution.

\begin{algorithm}[H]
\KwInput{$S = \{ s_{1}, s_{2} \cdots \}, T = \{ t_{1}, t_{2} \cdots \}, \tau, c$}
\KwOutput{$S^{'} = \{ s^{'}_{1}, s^{'}_{2} \cdots \}$}
$U \leftarrow \phi, t \leftarrow 0, S^{'} \leftarrow \phi$\\
\While{true}
{
  $I \leftarrow \phi$\\
  \For{each $s_{i}$ in $S$}
  {
      \uIf{$\mid s_{i} - U \mid \neq 0$ \textbf{and} $\frac{t+t_{i}}{c} \leq \tau$}
      {
	  $I.append(s_{i}$)
      }
  }
  \uIf{$\mid I \mid \neq 0$}
  {
      $k \leftarrow \textit{arg\,max}_{i \in I} \frac{\mid s_{i} - U \mid}{t_{i}}$\\    
      $U \leftarrow U \cup s_{k}$\\
      $t \leftarrow t+t_{i}$\\
      $S^{'}$.append($s_{k}$)\\
      $S$.delete($s_{k}$)\\
  }    
  \Else{\textbf{break}}
}
Output $S^{'}$
\caption{Picking optimal instance sets}
\label{heuristic}
\end{algorithm}


 Algorithm 2, based on \cite{SAJ}, takes as input the set of all instances, S, the set of corresponding execution times, $T$, number of cores, $c$ and the bound on total execution time, $\tau$. For brevity, an instance set is denoted $s_{i}$ and the corresponding execution time is denoted $t_{i}$. It is assumed that the instance sets have already been computed as explained in the previous section. During each iteration, Algorithm 2 selects a new instance set. The instance set, $s_{i_{j}}$, selected in iteration $j$ is that which maximizes $\frac{|s_{i}-\mathbb{U}_{j-1}|}{t_{i}}$ where $\mathbb{U}_{j} = \bigcup_{k=1}^{j} s_{i_{k}}$ and $\sum_{k=i_{1}}^{i_{j}} t_{k} < \tau$.

Algorithm 2 terminates when no instance set can contribute a fresh instance to the set of instances so far accumulated in $\mathbb{U}$ or if adding any instance set causes the total computation time to exceed $\tau$. An estimate of the success probability achievable with $c$ cores in time, $\tau$ is $\frac{|(\sum_{j=1}^{m}s_{i_j})|} {n}$ where $m$ is the total number of iterations executed by Algorithm 2. Given an arbitrary instance whose secret needs to be discovered, the output of Algorithm 2 is used as follows. Run Algorithm 1 repeatedly with input parameters corresponding to the ($p,e_{1},e_{2}$) parameters of instance sets $s_{i_{1}}$, $s_{i_{2}}$, $\cdots$ $s_{i_{m}}$.  

\begin{figure}[h]
\includegraphics[width=12cm, height=7cm]{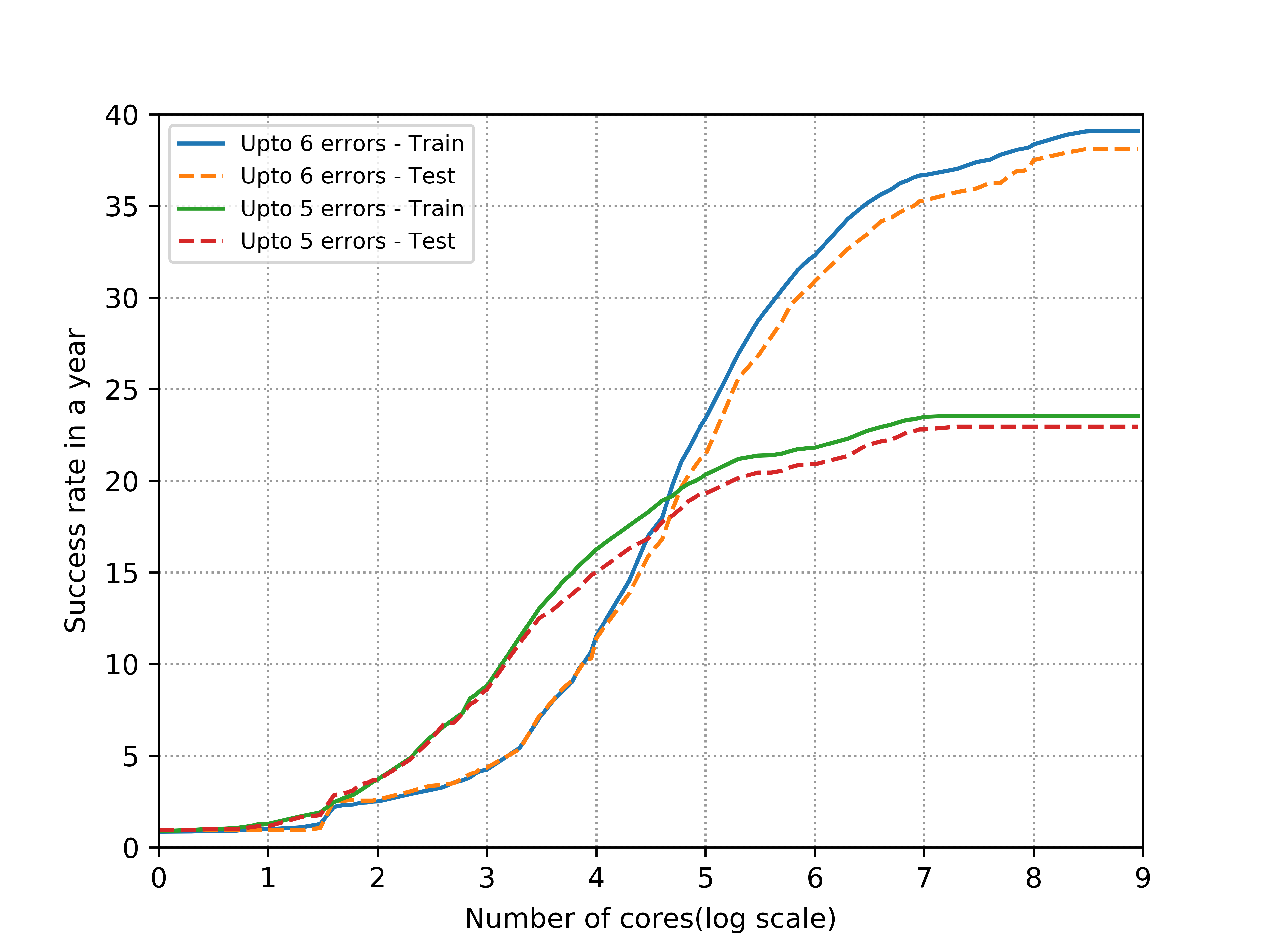}
\centering
\caption{Increase in success rate with increasing number of cores}
\label{success_cores}
\end{figure}

In the training phase, Algorithm 2 was run with instance sets derived from $10,000$ randomly generated instances. $\tau$ was fixed to be 1 year but the number of cores was varied. Two cases were considered - (i) $e_1 + e_2 = 5$ and (ii) $e_1 + e_2 = 6$. The success probability for each case with varying number of cores was estimated and plotted (Figure \ref{success_cores}). The success probability with $3000$ cores is $13\%$ and increases to over $15\%$ with $10,000$ cores for $e_1 + e_2 = 5$. This is considerably better compared to Strategy 2 ($6.1\%$ success probability).

Table \ref{PartitionTable} showed that the computational resources required to execute Algorithm 1 are substantially higher assuming $6$ rather than $5$ errors in the first $230$ bits of $u_s$. Hence the success probability with a smaller number of cores is higher for $e_1 + e_2 = 5$ compared to the case with $e_1 + e_2 = 6$. Beyond $50,000$ cores, the case of $6$ errors has much higher success probability. With $100,000$ cores,  the success probability is $23\%$ and increases to $33\%$ with $1000,000$ cores.

To test the efficacy of our approach, we generated $2000$ random test instances. Based on the results obtained by applying Algorithm 2 on the training data, we computed the average success probabilities of the test instances. The results as a function of number of cores is shown by dashed lines in Figure \ref{success_cores}. There is a very close match between the success rates obtained in the training and testing phases with a maximum discrepancy of around $4\%$.

%% file: Related_work.tex
\section{Related Work}

\cite{SDG13} attempt to learn some of the elements of $\textbf{u}$ and then use CVP to solve the reduced sub-instance. The following observation was made. If the elements of $\textbf{u}$ and $\textbf{A}$ are randomly chosen, then the average value of an element in $\textbf{c}_{1}$ would be $\frac{640}{4}= 160$. If the $i^{th}$ entry of $\textbf{c}_{1}$ is small and the Hamming weight of the $i^{th}$ row of $\textbf{T}$ is not especially low, then $\textbf{u}$ is likely to be $0$ in many of the positions corresponding to $1$'s in the $i^{th}$ row of $\textbf{T}$. Using this idea, some of the bits in $u$ may be guessed and CVP used to solve the reduced instance. However, \cite{SDG13} states that this method does not seem to be particularly effective beyond number of columns of $\textbf{T} = 400$.

\cite{b2} implemented parallel enumeration for the Bounded Distance Decoding (BDD) problem
and used it to solve Galbraith's first challenge. They solved Galbraith's first challenge using Ruhr-University's  ``Crypto Crunching Cluster" (C3) within 4.5 hours. However they did not report any results related to the solution of the second challenge.

Herold and May\cite{HM} studied the application of LP and ILP to obtain $\boldsymbol{u}$. They obtained results for $n$ = 256 and $m$ ranging from 400 to 640 for 1000 instances. The execution of a particular instance using ILP was aborted if it failed to obtain a solution within 10 seconds. The success probability dropped from 100\% at $m$ = 490 to 1\% at $m$ = 590.  Under certain mild assumptions, they also proved that the solution with LP relaxation for $m \leq 2n$ is unique. For any given instance they computed a score which quantifies the search space for the ILP 
. $2^{19}$ instances of GB-LWE were generated for $m = 640$. From this ensemble, 271 weak instances 
were identified. 16 of these were solved within half an hour each.  Since ILP is NP-hard and has, in general, exponential running time  they did not provide any time bound for solving an instance.


Herold and May's 
work was extended by  
\cite{SNDM} 
They presented an approach to classify an instance as 
easy, moderate or hard. Out of 100 easy instances from 1000 randomly generated instances they solved 5 instances in a day using 150 cores and 18 instances in 50 days using 3000 cores. They concluded that the increase in success rate could be achieved by exponential growth in the number of core-days.

%% file: Conclusion.tex
\section{Conclusion}

We addressed Galbraith's second challenge - recovery of the ephemeral key, \textbf{u} given a $256 \times 640$ matrix \textbf{T} and ciphertext $\textbf{c}_{1} = \textbf{T}\textbf{u}$. Our approach involved repeatedly guessing the first $230$ bits of \textbf{u} by modifying an initial guess based on the output produced by applying LP. Our first strategy was to create and solve the resulting sub-instance using CVP. The second strategy involved 2-step guessing of the $230$ bits before and after the second application of LP. This enhancement resulted in a larger number of instances with fewer number of errors in the initial guess thereby increasing success probability. Also, while there were a larger number of LR operations with Strategy 2, the reduced number of Babai NP computations was greatly reduced resulting in much lower overall computation time. With Strategy 1, we achieved a success rate of $9.4\%$ using about $50,000$ cores in 1 year while the success probability with Strategy 2 increased to $11.8\%$ using only $27,000$ cores in 1 year.

Strategy 3 makes repeated invocations to Algorithm 1 with different input parameters. The problem of learning the optimal input parameters is mapped to a variant of the Budgeted Maximum Coverage Problem. The parameters learned in the training phase surprisingly exhibited substantial diversity. Also, there was considerable variance in the execution times of the multiple runs of Algorithm 1 with the input parameters learned. This enhancement greatly increased the success probability to $16\%$ with approximately $27,000$ cores in 1 year. With $100,000$ cores the success probability touched $23\%$. One further avenue of investigation is the application of LP three or more times and the use of ILP or another method to solve the resulting sub-instance.

%% file: References.tex
\vspace{12pt}
\color{red}